\documentclass[12pt,draftclsnofoot,onecolumn]{IEEEtran} 
\usepackage{enumitem}
\usepackage{setspace}
\usepackage{graphicx}
\usepackage{subcaption}

\usepackage{color}
\usepackage{placeins}
\usepackage{float}

\usepackage[pdfencoding=auto,psdextra]{hyperref}
\usepackage{amsmath} 
\usepackage{amsthm}
\usepackage{amssymb}
\allowdisplaybreaks

\usepackage{xcolor}
\usepackage{pgf,tikz}
\usepackage{mathrsfs}
\usetikzlibrary{shapes,shadows,arrows}
\usepackage{marvosym}
\usepackage{multirow}
\usepackage{filecontents}
\usepackage[noadjust]{cite}
\usepackage{xfrac}
\newcommand{\Rom}[1]{\uppercase\expandafter{\romannumeral #1\relax}}
\newcommand{\rom}[1]{\lowercase\expandafter{\romannumeral #1\relax}}

\newcommand{\dd}{\mathrm{d}}

\newcommand{\bo}[1]{\boldsymbol{#1}}
\newcommand{\ql}{\textquoteleft }
\newcommand{\qr}{\textquoteright }
\newcommand{\Ql}{\textquotedblleft }
\newcommand{\Qr}{\textquotedblright }

\newcommand{\correct}[1]{{\color{black}#1}}
\newcommand{\modify}[1]{{\color{black}#1}}

\begin{document}
	\title{\modify{GLRT based Adaptive-Thresholding for CFAR-Detection of Pareto-Target in Pareto-Distributed Clutter} }
	
	\author{John~Bob~Gali,
		Priyadip~Ray,~\IEEEmembership{Member,~IEEE,}
		and~Goutam~Das,~\IEEEmembership{ Member,~IEEE}
		\thanks{John Bob Gali and Goutam Das are with the G.S. Sanyal School of Telecommunications, Indian Institute of Technology (IIT) Kharagpur, West Bengal 721302, India, e-mails: johnbob.neo@gmail.com, gdas@gssst.iitkgp.ac.in.}
		\thanks{Priyadip~Ray was with the G.S. Sanyal School of Telecommunications, Indian Institute of Technology (IIT) Kharagpur, West Bengal 721302, India, e-mail: priyadipr@gmail.com.}
		\vspace{-1cm}
	}
	\maketitle
	\begin{abstract}
After Pareto distribution has been validated for sea clutter returns in varied scenarios, some heuristics of adaptive-thresholding appeared in the literature for constant false alarm rate (CFAR) criteria. These schemes used the same adaptive-thresholding form that was originally derived for detecting \mbox{Swerling-I } (exponential) target in exponentially distributed clutter. Statistical procedures obtained under such idealistic assumptions would affect the detection performance when applied to newer target and clutter models, esp. heavy tail distributions like Pareto.

Further, in addition to the sea clutter returns, it has also been reported that Generalized Pareto distribution fits best for the measured Radar-cross-section (RCS) data of a SAAB aircraft. Therefore, in Radar application scenarios like Airborne Warning and Control System (AWACS), when both the target and clutter are Pareto distributed, we pose the detection problem as a two-sample, Pareto vs. Pareto composite hypothesis testing problem.

We address this problem by corroborating the binary hypothesis framework instead of the conventional way of tweaking the existing adaptive-thresholding CFAR detector. Whereby, for the composite case, considering no knowledge of both scale and shape parameters of Pareto distributed clutter, we derive the new adaptive-thresholding detector based on the generalized likelihood ratio test (GLRT) statistic. We further show that our proposed adaptive-thresholding detector has a CFAR property. We provide extensive simulation results to demonstrate the performance of the proposed detector.
	\end{abstract}

	\section{Introduction}	
	\IEEEPARstart{I}{n} automatic Radar Target detection, modeling and statistical signal processing play an important role in designing constant false alarm rate (CFAR) detection procedures. CFAR is an adaptive thresholding detection procedure intended to detect targets immersed in varying background clutter \cite{barton_radar_2004}.  
	
\modify{Contrasting a fixed-thresholding procedure (where the received signal is compared to the same detection threshold as the Radar sweeps across the resolution-cells), to maintain CFAR, the adaptive-thresholding procedure compares the signal received to a statistic that summarizes the local clutter strength's variability \cite{rohling1983radar}. In Radar terminology, cell-under-test (CUT) is a resolution cell on which we conduct the target detection and the reference-window, the cells surrounding the CUT (as in Figure \ref{fig:range}), are to estimate the clutter strength. For example, to detect exponential target (Swerling-I) in exponential clutter, the adaptive-thresholding procedure compares the CUT to a constant, multiplied by the mean of the  reference-window observations \cite{richards_principles_2010}. Whereby, the threshold adapts  to the relative mean of the  reference-window clutter strength as the Radar sweeps across the resolution-cells. Thus the name, cell-averaging CFAR (CA-CFAR), was promulgated and was proved optimal by posing the problem as exponential vs. exponential binary hypothesis testing \cite{gandhi_optimality_1994}.

	\begin{figure}[t!]
	\centering
	\includegraphics[width=0.75\textwidth,clip,trim=2cm 8cm 2cm 5cm]{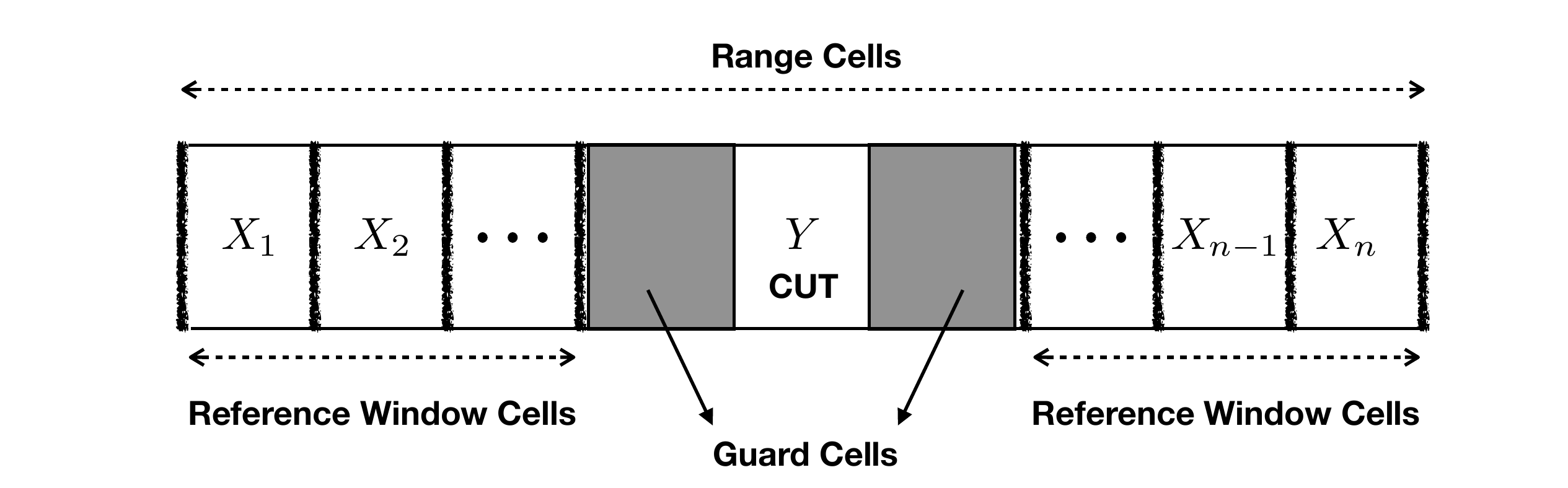}
	\caption{Typical range profile for target detection.}
	\label{fig:range}
\end{figure}
	Incidentally, the same form of adaptive thresholding (a constant multiplied by a statistic) is applied in different contexts for combating heterogeneous environments, multiple target scenarios, and even to the newer-clutter models under homogeneous environment to get the CFAR at the cost of detection performance \cite{richards_principles_2010, rohling1983radar}}.  Because of this, we see poor detection performance in detecting real-world targets like aircraft or ballistic targets \cite{johnston_target_1997}. In \cite{johnston_target_1997-1}, the authors reiterated the importance of accurate target fluctuating models, departing from the conventional Marcum and four Swerling models for improved detection performance.
	
		Primarily, detection performance (i.e., probability of detection $P_d$) is mainly attributed to the back-scattering from the target, often termed as radar cross-section (RCS) of the target \cite{wilson_probability_1972}. \correct{As RCS of simple targets, involves extremely complex formulation as a function of aspect angle, frequency, and polarization, there is a paradigm shift to the Statistical modeling of RCS \cite{richards_principles_2010}. So, the RCS within a single resolution cell is usually considered as a random variable from a specified distribution termed as a target-fluctuating-model \cite{richards_fundamentals_2005}.} Thus, accurate target models are essential for good detection performance.
		
			In the context of aircraft detection, it has been recently shown that generalized Pareto distribution best fits a SAAB aircraft's RCS, \correct{wherein the experimental setup, specifications, and characteristics of the aircraft navigation system along with the in-flight configurations are given in \cite{persson_radar_2017}.} Statistical models are widely accepted to capture these sensitive RCS fluctuations \cite{richards_principles_2010}, and we call any such Pareto distributed target-fluctuating-model as Pareto-Target (PT) \cite{Gali_2020}.
			
	On the other hand, modeling the spiky (heavy tail) behavior of sea clutter with Pareto distribution has gained much attention after it was validated for X-band high-resolution clutter intensity \cite{weinberg_noncoherent_2017,Rosenberg2019}. Furthermore, Pareto has been the forerunner for high-resolution sea clutter returns at low and high grazing angles, outperforming Log-normal, Weibull and $K$-distribution. However, it closely matches $KK-$distribution with five parameters \cite{farshchian_pareto_2010, weinberg_assessing_2011}. Thereby, many detection schemes \cite{Xue_2019,weinberg_2019_weberHdetector,weinberg_coherent_2011,weinberg_constant_2013,weinberg_assessing_2013} and constant false alarm rate (CFAR) detectors \cite{weinberg_constant_2013} were designed  for this clutter model.

		\correct{Lately, works in \cite{weinberg_examination_2014} mainly focused on fitting or adapting the solution that was obtained for Gaussian-intensity case (i.e., exponential vs. exponential hypothesis testing \cite{gandhi_optimality_1994}  where one considers both exponentially distributed target and clutter scenario) to the detection problem in Pareto clutter. In \cite{weinberg_general_2014}, assuming that the scale parameter is known and by exploiting the transformation of Pareto to the exponential distribution, geometric mean (GM)-CFAR detector was derived. In other words, it preserves the relation between the threshold and probability of false alarm ($p_{fa}$) as that was obtained for the Gaussian intensity case, i.e., the cell-averaging CFAR (CA-CFAR). In the subsequent work \cite{weinberg_constant_2014}, it is attributed that the CFAR process depends on the scale parameter for preserving the Gaussian \Ql threshold - $p_{fa}$\Qr \, relationship. Later on, in \cite{weinberg_construction_2017,weinberg_invariant_2017}, the scale parameter dependence on CFAR detection was rectified by employing the complete sufficient statistic. All of this existing literature focuses on the scenario where PT is not considered; wherein, there is an inherent loss in the detection performance. In contrast, our procedure provides an elegant solution, i.e., GLRT based CFAR detector, which we derive from the first principles by projecting the problem as a binary Composite hypothesis testing.}

			\correct{Further, in maritime surveillance and reconnaissance, an airborne early warning and control (AEW\&C) or AWACS (Airborne Warning and Control System) are popularized incorporating the radar picket in aerial warfare \cite{force_2019,defenceWorld}. In such aerial engagements, scenarios of detecting enemy aircraft over the sea clutter are quite common when viewed from a radar picket at higher altitudes. For example, patrol planes with onboard installed radars (AWACS) are often engaged to detect enemy aircraft hovering over the sea. In such scenarios, we address this problem of detecting a Pareto modeled aircraft target immersed in Pareto distributed clutter from the two-sample hypothesis testing framework.}
	
\modify{Thereby in this paper, for a preliminary case considering PT immersed in the Pareto clutter with known scale, the GLRT yields the same adaptive-thresholding, GM-CFAR, which is optimal in UMP sense \cite{Gali_2020}. We  further extend the work \cite{Gali_2020} by assuming a more realistic scenario, considering unknown shape \ql$\alpha$\qr{} and scale \ql$h$\qr{} parameters.} By making the scale parameter unknown, Pareto distribution (two-parameter) no longer belongs to the exponential family where the standard statistical procedures are readily available \cite{casella2002statistical}. To the best of our knowledge, we could not find any other literature on CFAR detection \correct{of PT in Pareto clutter} when both the parameters are unknown. 

	

	So, instead of trying to get a CFAR detector, we start from first principles by posing the problem as a composite binary hypothesis testing for the heavier tail, just as it is done for the exponential vs. exponential (i.e., detecting Swerling-\Rom{1} target in exponential clutter)  \cite{gandhi_optimality_1994}. \modify{ We then corroborate the new adaptive thresholding mechanism from the theoretical derivation of the GLRT test for Pareto vs. Pareto.}

	To summarize,  \begin{itemize}
			\item We pose the aircraft  detection problem as a two-sample hypothesis test for comparing the tail index of two-parameter Pareto distribution.
			\item We solve it systematically by GLRT approach and derive expressions for $p_d$ and $p_{fa} $.
			\item \modify{We verify the CFAR property of the derived adaptive-thresholding detector by theory and simulation.}
			\item We match analytical receiver operating characteristics (ROC) curves against simulated ROC curves.
	\end{itemize}  
	
We next provide rudiments about two-parameter Pareto distribution in section \Rom{2}, followed by system model, the formal statement of composite binary hypothesis testing problem. Next, we provide the detection procedures in a systematic manner of increasing complexity by relaxing assumptions on the knowledge of the parameters in sections \Rom{3} and \Rom{4}, respectively. We then give extensive simulation results validating to validate our proposed detector in section \Rom{5}. Section \Rom{6} concludes the paper.

	\section{Statistical Preliminaries}
	Pareto distribution is one among the power-law family with a negative exponent \cite{balakrishnan_primer_2004}. We use the notation $Y\sim Pa(\alpha,h)$ \space for a random variable $Y$ drawn from a Pareto distribution with shape $\alpha>0$ and scale $h>0$ parameters. Its cumulative distribution function (cdf) and probability density function (pdf) expressions are as follows:
	\begin{align}
	\label{eq:cdf}
	F_Y(y)&=1-\left(\frac{h}{y}\right)^\alpha, \quad y\geq h,  \\
	f_Y(y;\alpha,h)&=\frac{\alpha h^{\alpha}}{y_i^{\alpha+1}}I_{[h,\infty)}(y)
	\end{align}
	where the indicator function is defined as, $I_A(y)$ is one when $y\in A$, otherwise it is zero. In other words, support of $Y$ is parameterized interval $A=[h,\infty)$.
	
	\correct{For the squared amplitude or intensity observations in the range cells, we assume this two-parameter Pareto model $Pa(\alpha,h)$ \cite{weinberg_constant_2013}, where the shape parameter $\alpha$ dictates the fatness of the distribution tail and the scale $h$ defines the support set.}
	

%
		
	
	\section{System Model}
	Consider a Radar target detection problem in homogeneous background clutter wherein the squared amplitude or intensity observations in the range cells are modeled as Pareto distributed $Pa(\alpha,h)$ with the unknown shape $\alpha$ and scale $h$ parameter. The range cells comprise of reference window cells and a cell under test  (CUT) as shown in Fig. \ref{fig:range}. Reference window cells read the background clutter observations $X_1, X_2,\dotsc, X_n$, while the CUT reads a single observation $Y$, the backscattering either from the same background clutter or  from the target. Usually, CUT is isolated from the reference window cells by the several guard cells as shown in Fig. \ref{fig:range}. So, the CUT observation $Y$ is statistically independent of each of $X_i$'s. Even the reference window cells are sufficiently apart such that $X_i$'s are also independent. \correct{	So, for the discerning parameter, the tail-index $\alpha$, we choose a different notation $\rho$ for the CUT $Y,$  to distinguish it from that of window reference cells.}
	
	Now, we pose the problem as a two-sample test for comparing tail indices, with one sample lot $\bo{X}=(X_1, X_2,\dotsc, X_n)'$ consisting $n$ iid observations, each $X_i\sim Pa(\alpha,h)$, while the other sample lot has one observation $Y\sim Pa(\rho,h)$ on which the test is conducted. As lower values of shape parameter imply heavier tail, we say the target is present when $\rho<\alpha$, i.e., $\rho $ is restricted to $(0,\alpha)$, while $\alpha$ is unrestricted, allowing natural parametric space $(0,\infty)$. Also, in both the sample lots, the scale parameter $h$ takes the same value in $(0,\infty)$ and merely acts as a nuisance parameter. So, our two-sample hypothesis testing problem can be compactly stated as follows: 
	
	\textbf{Problem Statement:}
	\emph{
		Let $(\bo{x},y)$ be the realization of two independent random sample lots $(\bo{X},Y) $, drawn from $Pa(\alpha,h)$ and $Pa(\rho,h)$ respectively. Our problem is to find GLRT test for the hypotheses
		\begin{equation}
		\label{eq:problem 1}
		\begin{aligned}
		H_0 &:\; \rho = \alpha \quad \\ \text{vs.} \quad
		H_1 &:\; \rho < \alpha,
		\end{aligned}
		\end{equation}
		\begin{enumerate}[label=\textnormal{case (\alph*)},align=left]
			\item \hspace{-4pt}\textnormal{:} when $\alpha$ is unknown and $h$ is known\label{itm:1};
			\item \hspace{-4pt}\textnormal{:} when both $\alpha$ and $h$ are unknown\label{itm:2}.
		\end{enumerate}
	}
	Before addressing the above cases, (a) and (b), we shall study a simple scenario, though unrealistic, but helps us in getting  upper bounds on the detection performance. We call this idealized scenario \ql simple vs. composite,\qr{} based on the specification of parameters in the probabilistic model, as described below.
	\subsection{ \ql Simple vs. Composite\qr{}}	
\begin{figure}[H]
		\centering 
		\includegraphics[width=0.75\textwidth]{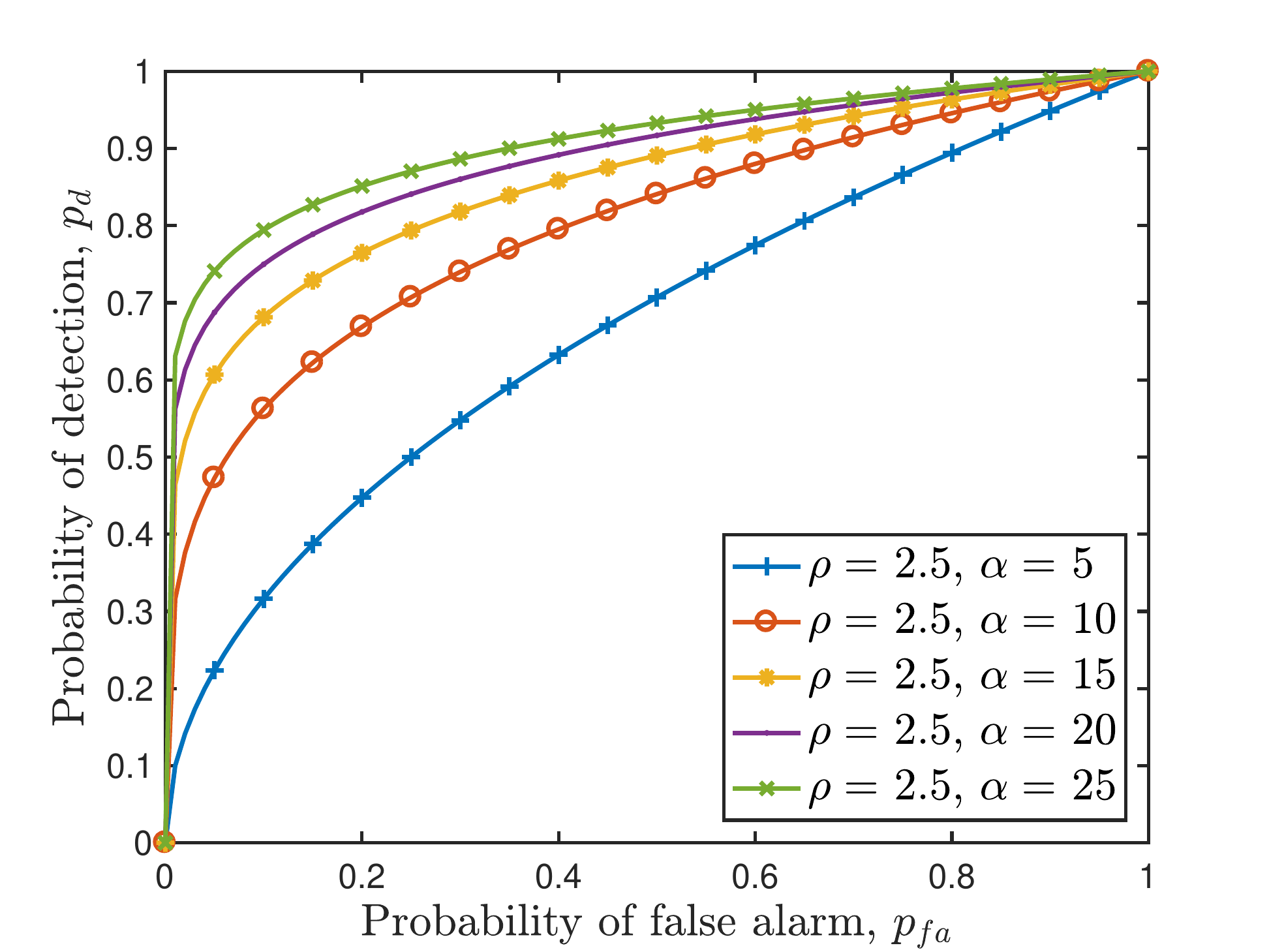}
		\caption{Family of ROC curves for target detection in known clutter parameters.}
		\label{fig:FamilyOfRoc}
	\end{figure}

	We assume perfect knowledge of the clutter statistics, i.e., $ \alpha$ and $h$. So the reference window observations become irrelevant here. Then clearly, the null hypothesis $H_0$ is \ql simple,\qr{} as $Y\sim Pa(\rho,h) \text{ with }\rho=\alpha$ is completely specified when no target is present, and the alternate hypothesis $H_1$ is  \ql composite,\qr{} as $\rho\in (0,\alpha)$.  The usual strategy for this scenario is to pretend that we know the exact value of $\rho$ (thereby presuming $H_1$ simple), and derive the Neyman Pearson (NP) test. Then, if we can find the test statistic and the threshold without utilizing the knowledge of the unknown parameter $\rho$, then the test so derived is optimal and can be used as an upper bound (clairvoyant detector) for detection performance \cite{kay_fundamentals_1998}. Therefore by NP-lemma, the likelihood ratio test (LRT) is
	\begin{equation}
	\lambda (y) =\frac{f_Y(y;H_1)}{f_Y(y;H_0)} \overset{H_1}{\underset{H_0}{\gtrless}}\gamma, 
	\end{equation}
	and after some simplifications we get 
	\begin{equation}
	y\overset{H_1}{\underset{H_0}{\gtrless}} h\left(\frac{\alpha \gamma}{\rho}\right)^{1/\left(\alpha - \rho\right)}=\gamma_{th} \;\text{(say).}
	\end{equation}
	Even though the threshold $\gamma_{th}$ is dependent on unknown parameter $\rho$, as the test statistic $Y$ is independent of $\rho$, we can choose a threshold for any predetermined significance level or the probability of false alarm $p_{fa}$. Hence, 
	\begin{equation}
			\begin{aligned}
	 	p_{fa}&=\Pr\left(Y>\gamma_{th};H_0\right)\\
		&=\left(\frac{h}{\gamma_{th}}\right)^\alpha,
		\end{aligned}
	\end{equation}
	since it is  complimentary to the cdf \eqref{eq:cdf}, so that
	\begin{equation}
	\label{eq:threshold pfa}
	\gamma_{th}=\frac{h}{{p_{fa}}^{1/\alpha}}.
	\end{equation}
	 As the threshold $\gamma_{th}$ and $p_{fa}$ relation \eqref{eq:threshold pfa} doesn't involve the unknown parameter $\rho$, we have the optimal test in NP sense.
	Similarly, the probability of detection is given by 
	\begin{equation}
		\begin{aligned}
		p_{d}&=\Pr\left(Y>\gamma_{th};H_1\right)\\ 
		&=\left(\frac{h}{\gamma_{th}}\right)^\rho,
		\end{aligned}
	\end{equation}
	and is dependent on the unknown $\rho$. Further, after substituting $\gamma_{th}$ from equation \eqref{eq:threshold pfa} and rearranging, we get
	\begin{equation}
	p_{d}=(p_{fa})^{\rho/\alpha}.
	\label{eq:rocRelation}
	\end{equation}
	Therefore, the ROC curves directly follow from  \eqref{eq:rocRelation} for varied clutter tail index, $\alpha$ as shown in Fig. \ref{fig:FamilyOfRoc}. These curves can be used as an upper-bound when we relax the assumptions on knowledge of clutter parameters.

	\subsection{ \ql Composite vs. Composite\qr{}} When we relax the assumptions on clutter parameters, clearly, the null hypothesis becomes composite and choosing a threshold for a particular $p_{fa}$ requires the knowledge of parameters under null hypotheses. So, by GLRT approach, we circumvent the problem by using maximum likelihood (ML) estimates of the unknown parameters. Usually, for any two sample $(\bo{x,y})$ hypothesis testing, $H_0: \bo{\theta} \in \bo{\Theta_0}$ vs. $H_1: \bo{\theta }\in \bo{\Theta} \setminus \bo{\Theta_0} (\text{say }\bo{\Theta_1})$, by GLRT, the critical region (i.e., rejecting $H_0$, or deciding $H_1$) is given as
	\begin{align}\label{eq:criticalRegion}
	\mathcal{R} &=\{(\bo{x,y}): \lambda(\bo{x,y})<\gamma\} \\ \text{where}\quad
	\lambda(\bo{x,y}) &=\frac{\sup_{\bo{\theta} \in \bo{\Theta}_0} L(\bo{\theta},\bo{x,y})}{\sup_{\bo{\theta} \in \bo{\Theta}}L(\bo{\theta},\bo{x,y})},
	\end{align}
	is the generalized likelihood ratio (LR) which ranges between zero and one. So, for a particular value of $p_{fa}$, we choose $\gamma$ such that the size of the test  $\sup_{\bo{\theta} \in \bo{\Theta}_0} \Pr(\lambda(\bo{X,Y}) < \gamma)=p_{fa}$. Thus, for our two sample $(\bo{x},y)$ hypothesis test \eqref{eq:problem 1}, after replacing the unknown parameters with their respective restricted and whole parametric ML estimates, $\hat{\bo{\theta}}_0$ and $\hat{\bo{\theta}}$, the GLRT statistic is given by
	\begin{equation}
	\lambda(\bo{x},y) = \frac{L(\hat{\bo{\theta}}_0,\bo{x},y)}{L(\hat{\bo{\theta}},\bo{x},y)} \overset{H_1}{\underset{H_0}{\lessgtr}} \gamma.
	\end{equation}
	We address the problem \eqref{eq:problem 1} by this GLRT given above, for each case separately in the following section. 
	
	\section{Solution by GLRT for \ql Composite vs. Composite\qr{}}	
	\subsection{\ref{itm:1}\textnormal{:} When $\alpha$ is unknown, and $h$ is known.}
	In this case, the parametric spaces under null, alternate, and whole parameter spaces are:
	\begin{equation}
	\label{eq:parSpace}
	\begin{aligned}
	{\Theta}_0&=\{\alpha:\rho = \alpha,\;\alpha\in(0,\infty)\}, \\
	\boldsymbol{\Theta}_1&=\{(\alpha,\rho)':\;\rho \in (0,\alpha),\;\alpha\in(0,\infty)\},
	\\ \text{and }	\boldsymbol{\Theta}&=\{(\alpha,\rho)':\;\rho \in(0,\alpha],\;\alpha\in(0,\infty)\}.
	\end{aligned}
	\end{equation}
	As the densities of the $two$ lots $(Y,\bo{X})$,  $\bo{X}$ with each $X_i$, and $Y$ are independent, the likelihood $L$, for the whole parameter space $\bo{\theta} \in \bo{\Theta}$ is given as the product of densities, i.e.,
	\begin{equation}
	\label{eq:likelihood}
	L({\bo{\theta}},\bo{x},y)=\frac{\rho h^{\rho}}{y^{\rho +1}}\frac{\alpha^n h^{\alpha n}}{\left(\prod_{i=1}^{n} x_i\right)^{\alpha +1}} 
	\end{equation}
	where $\bo{\theta}=(\alpha,\rho)'$ is a column vector of unknown parameters.
	Here, we absorbed the indicator function for simplicity as $h$ is known, and the support of the random variable is understood as $ [h,\infty)$. As the logarithm is monotonically increasing function, we consider logarithm of the likelihood in finding the ML estimates. Therefore, by taking logarithm on both sides of  \eqref{eq:likelihood},  the log-likelihood function, $l(\bo{\theta})$ is
	\begin{multline}
	\label{eq:loglikelihood}
	l(\bo{\theta})=\ln L(\bo{{\theta}},\bo{x},y)= n\ln \alpha +\ln \rho + \\(n\alpha + \rho)\ln h - (\rho + 1)\ln y -(\alpha +1)\ln \left(\prod_{1}^{n} x_i\right).
	\end{multline}
	\subsubsection{Under ${\Theta}_0$}Firstly, we substitute $\rho=\alpha$ in $l(\theta)$ in \eqref{eq:loglikelihood}, making it a single variable function. Then, for maximizing $l$ with respect (w.r.) to $\alpha$ on $(0,\infty)$, as given under ${\Theta_0}$ in  \eqref{eq:parSpace}, we make the first derivative vanish. I.e., setting 
	\begin{equation}
			\frac{\mathrm{d} l}{\mathrm{d} \alpha}= \frac{n+1}{\alpha} +(n+1)\ln h-\ln \left(\prod_{i=1}^{n}x_i\right)-\ln y=0,
	\end{equation}
	we get a stationary point at
	\begin{align}
	\alpha= \frac{n+1}{\ln \left(\frac{y}{h}\prod_{i=1}^{n} \left[\frac{x_i}{h}\right]\right)}=\hat{\alpha}_{\Theta_0}\;(\text{say}.
	\label{eq:alphahat0}
	\end{align}
	Now, by the second derivative test, we see that the concavity,  $\left.\frac{\mathrm{d}^2 l}{\mathrm{d} \alpha^2}\right|_{\alpha=\hat{\alpha}_{\bo{\Theta}_0}}$ is negative. Since we have one stationary point, the absolute maximum of $l$ is attained at \ql$\hat{\alpha}_{\Theta_0},$\qr{} which we call ML estimate of $\alpha$ under $\Theta_0$. 
	\subsubsection{Under $\bo{\Theta}$}
	Here, the log-likelihood is a function of two variables, $ \bo{\theta} =(\alpha, \, \rho)' \in \bo{\Theta}$ as  given in \eqref{eq:parSpace}. \correct{The parameter space of $\Theta$ is constrained by the relations $\rho\in(0,\alpha],\alpha\in(0,\infty)$. We seek to maximize the log-likelihood under these constraints. It is evident that for $\rho=\alpha$, we shall obtain the same solutions as that under ${\Theta}_0$. Thus, we are left with the case when $0<\rho<\alpha$.  By setting the first-order partial derivatives 
		\begin{equation}
		\begin{aligned}
		\frac{\partial l}{\partial \alpha}&=\frac{n}{\alpha} + n\ln h -\ln \left(\prod_{i=1}^{n}x_i\right)=0,\\
		\frac{\partial l}{\partial \rho}&=\frac{1}{\rho}+\ln h-\ln (y)=0,
		\end{aligned}
		\end{equation}
		we get an unrestricted stationary point $(\hat{\alpha},\hat{\rho})$, given as
		\begin{equation}
		\begin{aligned}
		\label{eq:mlEstimates}
		\hat{\alpha}&=\frac{n}{\ln (\prod_{i=1}^{n}\frac{x_i}{h})}=\frac{n}{\Lambda(\bo{x})}, \\
		\hat{\rho}&=\frac{1}{\ln (\frac{y}{h})}=\frac{1}{\Lambda(y)}.
		\end{aligned}
		\end{equation}
		Next, due to the constraint domain of parameter space, we analyze the behavior more closely. Firstly, we observe that the obtained value of $\alpha$ ($\hat{\alpha}$) satisfies the condition $\alpha >0$ and does not depend on $\rho$. When  $\alpha <\frac{n}{\Lambda(\bo{x})}$, $\frac{\partial l}{\partial\alpha}>0$ (likelihood function increases), while for $\alpha >\frac{n}{\Lambda(\bo{x})}$, $\frac{\partial l}{\partial\alpha}<0$ (likelihood function decreases). This implies that $\hat{\alpha}$ maximizes the likelihood for any given $\rho$. Similar argument hold true for $\hat{\rho}$, and hence the 
		pair ($\hat{\alpha},\hat{\rho}$) will be the optimal pair for the unrestricted case. The condition $\rho<\alpha$ on the constraint set will impose the condition that $\hat{\rho}<\hat{\alpha}$ implies $ \frac{1}{\Lambda(y)}<\frac{n}{\Lambda(x)} $. If this condition is violated, we understand that the likelihood increases for all values of $\rho<\hat{\rho}$, while the constraint $\rho\leq\alpha$ would allow $\rho$ to only reach till $\alpha$. Thus, the solution will always be at the boundary i.e., when $\rho=\alpha$. This solution should be the same as that found under $\Theta_0$ in \eqref{eq:alphahat0}. Therefore, the MLE is given as:
		\begin{align}
		(\hat{\alpha},\hat{\rho})&={\begin{cases}
			\big(\frac{n}{\Lambda(x)}, \frac{1}{\Lambda(y)}\big) &\text{if   } \frac{1}{\Lambda(y)}<  \frac{n}{\Lambda(x)},  \\
			 \left(\hat{\alpha}_{\Theta_0}, \hat{\alpha}_{\Theta_0}\right)  &\text{otherwise.} 
			\end{cases}}  		
		\end{align}}
	When $ \frac{1}{\Lambda(y)} \geq\frac{n}{\Lambda(x)} $, the LR becomes one and we always accept $H_0$. So the critical region is mainly dictated when $\frac{1}{\Lambda(y)} < \frac{n}{\Lambda(x)}$. Therefore, the LR becomes,
	\begin{align}
	\label{eq:LR}
	\lambda(\bo{x},y)&=\frac{\hat{\alpha}_{\Theta_0}^{n+1} h^{{\hat{\alpha}_{\Theta_0}}(n+1)} y^{-(\hat{\alpha}_{\Theta_0}+1)} {\left(\prod_{i=1}^{n} x_i\right)^{-(\hat{\alpha}_{\Theta_0} +1)}} }
	{\hat{\rho}_{\Theta} \hat{\alpha}_{\Theta}^{n}h^{(\hat{\alpha}_{\Theta}n+\hat{\rho}_{\Theta})} y^{-(\hat{\rho}_{\Theta}+1)}     {\left(\prod_{i=1}^{n} x_i\right)^{-(\hat{\alpha}_{\Theta} +1)} }        }  \\
	&={(n+1)^{n+1}} \frac{n\frac{\Lambda(y)}{\Lambda(\bo{x})}}{\left((n\frac{\Lambda(y)}{\Lambda(\bo{x})} +n)\right)^{n+1} }.\label{eq:LRsimplified}
	\end{align}
	After substituting the MLEs in \eqref{eq:LR}, arriving at \eqref{eq:LRsimplified} is a non-trivial step, and we give the simplification in the appendix. From \eqref{eq:LRsimplified}, we can see that the LR is dependent on the data through $ n\frac{\Lambda(y)}{\Lambda(\bo{x})}$. So, letting $u=n\frac{\Lambda(y)}{\Lambda(\bo{x})}$, we have LR as a decreasing function of $u$ because
	\begin{equation}
	\frac{\mathrm{d}\lambda}{\mathrm{d}u} =\frac{n(n+1)^{n+1}(1-u)}{(n+u)^{n+2}} < 0,
	\end{equation}
	since from the critical region condition $\frac{1}{\Lambda(y)}<\frac{n}{\Lambda(x)}$, we have  $u>1$. Therefore, the critical region $\lambda(u)<\gamma$ is equivalent to $u>\gamma_1$, where $\gamma_1=\lambda^{-1}(\gamma)$. So, for a given value of $p_{fa}$, we can choose $\gamma_1$ from the size condition
	\begin{align}\label{eq:sizeSimplification1}
	p_{fa}&=\sup_{{\theta_0} \in {\Theta_0}}\Pr(u>\gamma_1)\nonumber\\ 
	&=\sup_{{\alpha} \in {(0,\infty)}}\Pr\left(\frac{\ln\left(\frac{Y}{h}\right)}{\frac{1}{n}\sum_{i=1}^{n}\ln\left(\frac{X_i}{h}\right)}>\gamma_{1}\right)\nonumber\\
	&=\sup_{{\alpha} \in {(0,\infty)}}\Pr\left( \frac{B}{C}>\gamma_{1} \right).
	\end{align}
	Here, $B=\ln\left(\frac{Y}{h}\right)\sim \mathrm{Exp}(\alpha)$ exponentially distributed with rate $\alpha$ and $C={\frac{1}{n}\sum_{i=1}^{n}\ln\left(\frac{X_i}{h}\right)}\sim \mathrm{Gamma}\left(n,\frac{1}{n\alpha}\right)$, standard gamma distributed with shape parameter $n$, and scale parameter $\frac{1}{n\alpha}$. This is because of the transformation of random variables, i.e., the logarithm of scaled Pareto distributed is exponential, and the sum of exponential distributed random variables is gamma-distributed. Whereby, after simplifying the  \eqref{eq:sizeSimplification1}  (given in appendix), the relation between $p_{fa}$ and threshold $\gamma_{1}$ is
	\begin{equation}\label{eq:threshold-pfa1}
	\begin{aligned}
	p_{fa}&=\left[1+\frac{\gamma_{1}^{}}{n}\right]^{-n}\\
	\text{or  }\gamma_{1}&= n\left(p_{fa}^{-1/n}-1\right),
	\end{aligned}
	\end{equation} 
	which is independent of unknown $\alpha$, the shape parameter of background clutter. So, our test statistic $u$ has CFAR property, and the GLRT is 
	\begin{align}\label{eq:glrtStatistic}
	n\frac{\Lambda(y)}{\Lambda(\bo{x}) } \overset{H_1}{\underset{H_0}{\gtrless}} \gamma_1.
	\end{align}
	
	Similarly for probability of detection $p_d=\Pr\left( \frac{B}{C}>\gamma_{1} ;H_1\right)$; under $H_1$, $B\sim \mathrm{Exp}(\rho)$ and $C$ remains $\mathrm{Gamma}\left(n,\frac{1}{n\alpha}\right)$. So following the simplification given in the appendix, we arrive at
	\begin{align}\label{eq:pd1}
	p_d=\left(1+\frac{\rho \gamma_1}{\alpha n}\right)^{-n}.
	\end{align}
	\subsection{\ref{itm:2} When both $\alpha$, and $h$ are unknown.}
	Here, with the introduction of $h$ as an unknown parameter, Pareto distribution no longer belongs to the regular class of exponential family as the density is given in terms of an indicator function of parameterized interval  $I_{[h,\infty)}(y)$. We can see this in the following expressions for the densities of $two$ lots, $\bo{X}$ with each $X_i\sim \mathcal{P}_a(\alpha,h)$ iid, for $i= 1,2, \dotsc, n$ and $Y\sim \mathcal{P}_a(\rho,h)$ given as
	\begin{equation}
		\begin{aligned}
		&f_{\bo{X}}(\bo{x};\alpha,h) =	\resizebox{0.75\hsize}{!}{$ \prod\limits_{i=1}^{n}\frac{\alpha h^{\alpha}}{x_i^{\alpha+1}}I_{[h,\infty)}(x_i) = \frac{\alpha^n h^{\alpha n}}{(\prod_{1}^{n} x_i)^{\alpha+1}}I_{[h,\infty)}(x_{(1)})$}, \\
		&\text{and }
		f_Y(y;\rho,h) = \frac{\rho h^{\rho}}{y^{\rho+1}}I_{[h,\infty)}(y)
		\end{aligned} 
	\end{equation}
	respectively. Here, $x_{(1)} = min (x_1,x_2,...,x_n)$, and the indicator function $I_{[h,\infty)}(z)$ is one when $z\in [h,\infty)$, zero otherwise. Therefore, the likelihood function $L(\boldsymbol{\theta},\bo{x},y)$ of the sample lots $(\bo{x},y)$ can be expressed as 
	\begin{equation}\label{eq:Likelihood}
	L({\bo{\theta}},\bo{x},y)=\frac{\rho h^{\rho}}{y^{\rho +1}}\frac{\alpha^n h^{\alpha n}}{\left(\prod_{1}^{n} x_i\right)^{\alpha +1}} I_{[h,\infty)}(min(y,x_{(1)})).
	\end{equation}
	Here $\boldsymbol{\theta}=(\alpha,\rho,h)'$ is a column vector, and the null, alternate and whole parameter spaces are as follows respectively: 
	\begin{equation}\label{eq:parSpace1}
	\begin{aligned}
	\boldsymbol{\Theta}_0&=\{(\alpha,\rho,h)':\;\rho = \alpha,\;\alpha\in(0,\infty),\;h\in(0,\infty)\} \\
	&=\{(\alpha,h)':\;\alpha\in(0,\infty),\;h\in(0,\infty)\},\\
	\boldsymbol{\Theta}_1&=\{(\alpha,\rho,h)':\;\rho \in (0,\alpha),\;\alpha\in(0,\infty),\;h\in(0,\infty)\}, \\ \text{and }
	\boldsymbol{\Theta}&=\{(\alpha,\rho,h)':\;\rho \in(0,\alpha],\;\alpha\in(0,\infty),\;h\in(0,\infty)\}.
	\end{aligned}
	\end{equation}
	Here, $h$ is unrestricted in the above parameter spaces \eqref{eq:parSpace1}, and it is acting as a nuisance parameter. Clearly, while fixing other variables in \eqref{eq:Likelihood}, we see that $L$ is monotonically increasing function of $h$ and we can deduce from the indicator function that $h$ takes its value from $(0,\min(y,x_{(1)})]$. So, the supremum is attained on the boundary, at $\min(y,x_{(1)}) = \hat{h}$ (say). Here we don't use the subscript to denote which parameter space $h$ belongs, as it is common to all parameter spaces. Now we look for MLE separately for the remaining unknown parameters $\alpha$ and $\rho$.
	\subsubsection{Under $\bo{\Theta}_0$\:}
	Here, we first replace $\rho$ with $\alpha$, just as the previous case in the log-likelihood \eqref{eq:loglikelihood}. After replacing $h$ with $\hat{h}$, by the first derivative test, supremum of $l$ w.r. to $\alpha$ is attained at
	\begin{align}
	\alpha= \hat{\alpha}_{\Theta_0}&= \frac{n+1}{\ln \left(\frac{y}{\hat{h}}\prod_{i=1}^{n} \left[\frac{x_i}{\hat{h}}\right]\right)}\nonumber\\
	&= \frac{n+1}{\Lambda(y)+\Lambda(x)},
	\end{align}
	where, by abusing the notation for $\Lambda(.)$ to include $\hat{h}$ in lieu of $h$, we now have
	\begin{equation}\label{eq:modifed Lamda}
	\begin{aligned}
	\Lambda(y)&=\ln\left(\frac{y}{\min(y,x_{(1)})}\right) \\ \text{ and }\Lambda(x)&=\ln\left({\prod_{i=1}^{n}\left(\frac{x_i}{\min(y,x_{(1)})}\right)}\right).
	\end{aligned}
	\end{equation}
	Note that we modified the notation $\Lambda(.)$ which was first introduced in \ref{itm:1}, in order to have the similar expression for test statistic and for ease of simplification.
	\subsubsection{Under $\bo{\Theta}$} 
	Incidentally, after replacing $h$ with $\hat{h}$ with $\Lambda(.) $ modified as in \eqref{eq:modifed Lamda}, following  similar steps as in the previous \ref{itm:1} under $\bo{\Theta}$, the MLEs of $\alpha$ and $\rho$ are exactly the same expressions. Further, following the same analysis and the same simplification steps, we could arrive at the same expression for the test statistic \eqref{eq:glrtStatistic}. Therefore, with the new modified $\Lambda(.)$, the critical region is
	\begin{equation}
	\label{eq:final test glrt}
	LR=\frac{\ln\left(\frac{y}{min(x_{(1)},y)}\right)}{\frac{1}{n}\sum_{i=1}^{n}\ln\left(\frac{x_i}{min(x_{(1)},y))}\right)}{>}\gamma, 
	\end{equation}
	for some $\gamma > 1$. Further, considering the numerator
	\begin{align}
	\ln\left(\frac{y}{min(x_{(1)},y)}\right)&={\begin{cases}
		\ln\left(\frac{y}{x_{(1)}} \right)&\text{if   } y > x_{(1) }\\	0 &\text{if  }  y\le x_{(1)},
		\end{cases}}  		
	\end{align}	
	we accept $H_{0}$, whenever $LR =0$ i.e., when $y\leq x_{(1)}$. So, the critical region is considered when $y>x_{(1)}$, such that the GLRT is now simplified to
	\begin{equation}\label{eq:glrt2}
	\frac{\ln\left(\frac{y}{x_{(1)}}\right)}{\frac{1}{n}\sum_{i=1}^{n}\ln\left(\frac{x_i}{x_{(1)}}\right)} \overset{H_1}{\underset{H_0}{\gtrless}} \gamma.
	\end{equation}
	Therefore, for a given significance level $p_{fa}$, the size condition is 
	\begin{align}\label{eq:sizeCondition}
	p_{fa}=\sup_{{(\alpha,h)} \in \Theta_0}\Pr\left(\frac{\ln\left(\frac{Y}{X_{(1)}}\right)}{\frac{1}{n}\sum_{i=1}^{n}\ln\left(\frac{X_i}{X_{(1)}}\right)}>\gamma\right). 
	\end{align}
	From theorem 3 of Malik's work \cite{malik_estimation_1970}, we have 
	\begin{equation} D=\frac{1}{n}\sum_{i=1}^{N}\log\left(\frac{X_i}{X_{(1)}}\right) \sim \mathrm{Gamma} \left(n-1,\frac{1}{\alpha n}\right). 
	\end{equation}
	Further in \cite{malik_estimation_1970}, he also proved that 
	$X_{(1)}$ independent of $D$. Also, it is easy to see that $X_{(1)} \sim $$P_a(n\alpha,h)$ \cite{balakrishnan_primer_2004}. Therefore, the size condition \eqref{eq:sizeCondition} becomes
	\begin{align}
	p_{fa} =\sup_{{(\alpha,h)} \in \Theta_0}\Pr\left(\frac{B-A}{D}>\gamma\right)
	\end{align}
	where, $A=\ln\left(\frac{X_{(1)}}{h}\right) \sim \mathrm{Exp} (n\alpha)$ and  $B=\ln\left(\frac{Y}{h}\right)\sim \mathrm{Exp}(\alpha)$. Let's denote $G=B-A$, difference of exponential distributed random variables whose density function can be derived as 
	\begin{equation}\label{eq:densityG}
	f_{G}(g)={\begin{cases}
		\frac{n}{(n+1)}\alpha e^{-\alpha g}&\text{if   } g>0\\	 \frac{n}{(n+1)}\alpha e^{n\alpha g}&\text{if   } g<0.
		\end{cases}} 
	\end{equation}
	Therefore, on further simplifying the size condition expression 
	\begin{equation}\label{eq:size2Simplificaiton}
	p_{fa}=\sup_{{(\alpha,h)} \in \Theta_0}\Pr\left(\frac{G}{D}>\gamma\right)
	\end{equation} in the appendix, the relation between $p_{fa}$ and threshold is 
	\begin{equation}
	\begin{aligned}\label{eq:threshold-pfa2}
	p_{fa}&=\frac{n}{n+1}\left[1+\frac{\gamma}{n}\right]^{-(n-1)}\\
	\text{or }	\gamma&= n\left(\left[\frac{n+1}{n}p_{fa}\right]^{1/(1-n)}-1\right),
	\end{aligned}
	\end{equation}
	which is independent of both the unknowns, $\alpha$ and $h$ parameters of the background clutter.  So our test statistic in \eqref{eq:glrt2} has CFAR property.
	
	For evaluating $p_d$, under $H_1$ the pdf of $G$ for $g>0$ gets modified to $\frac{\rho n\alpha}{(\rho+n\alpha)}e^{-\rho g}$ and $D$ remains same $\mathrm{Gamma} \left(n-1,\frac{1}{\alpha n}\right)$. So by following similar steps as in the previous \ref{itm:1}, we can arrive at 
	\begin{equation}\label{eq:pd2}
	p_d=\frac{n\alpha}{\rho+n\alpha}\left[1+\frac{\rho\gamma}{n\alpha}\right]^{-(n-1)}.
	\end{equation}
	
	\correct{	Thus for both \ref{itm:1} and \ref{itm:2}, for the test statistic so derived has no parameters in $p_{fa}$ expressions, but as expected, has both the clutter and target model parameters in $p_d$ expression. In other words, detection performance depends on the relative measure of the target to the clutter tail indices. In the next section, we validate our expressions with the Monte Carlo simulations. } 
\begin{figure*}[!t]
	\centering 
	\begin{subfigure}[t]{0.48\textwidth}
		\includegraphics[width=1\textwidth]{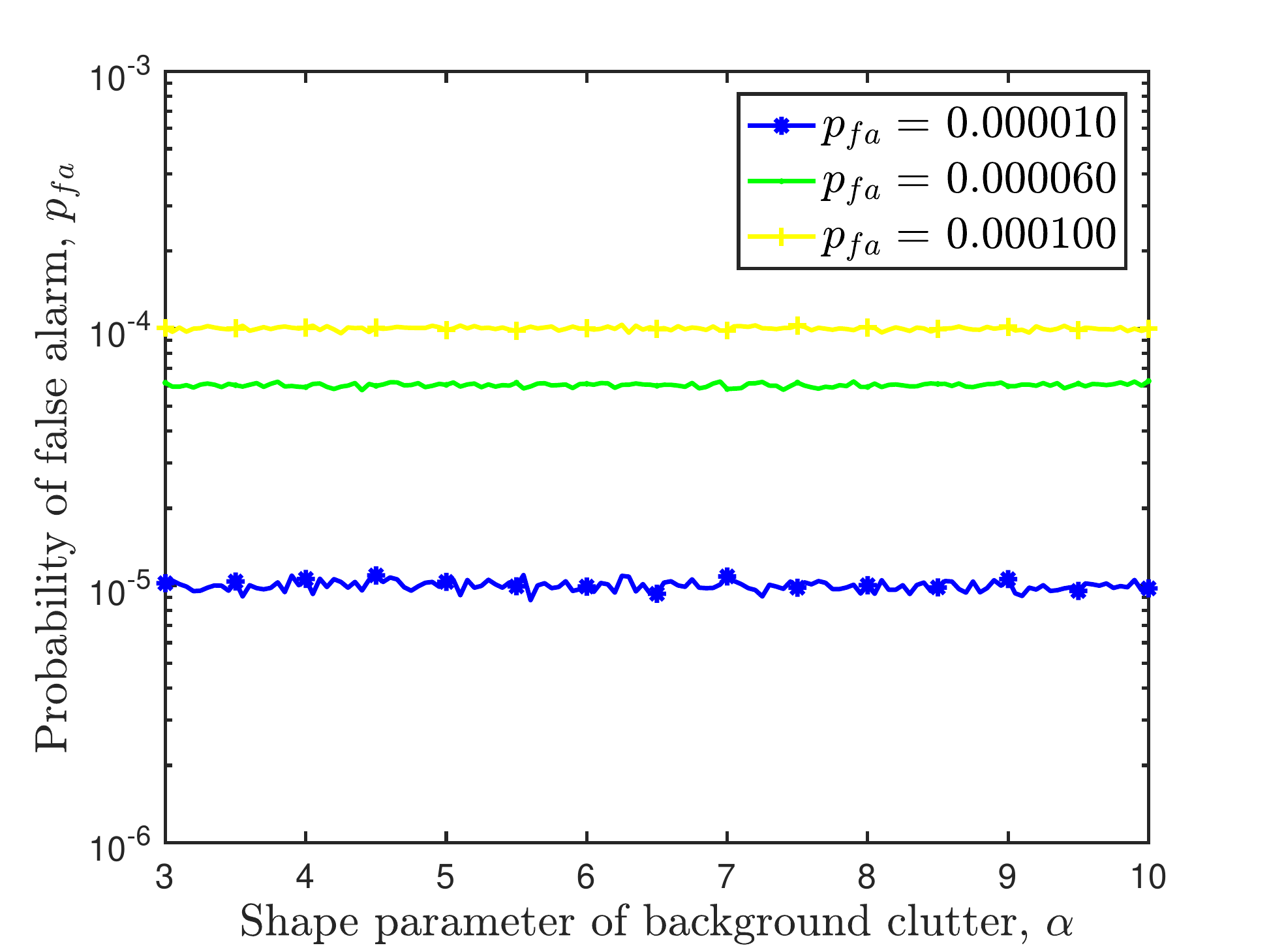}
		\caption{ GLRT-CFAR for varied unknown $\alpha$ and known h.}
		\label{fig:glrt_cfar1}
	\end{subfigure}
	\begin{subfigure}[t]{0.48\textwidth}
		\includegraphics[width=1\textwidth]{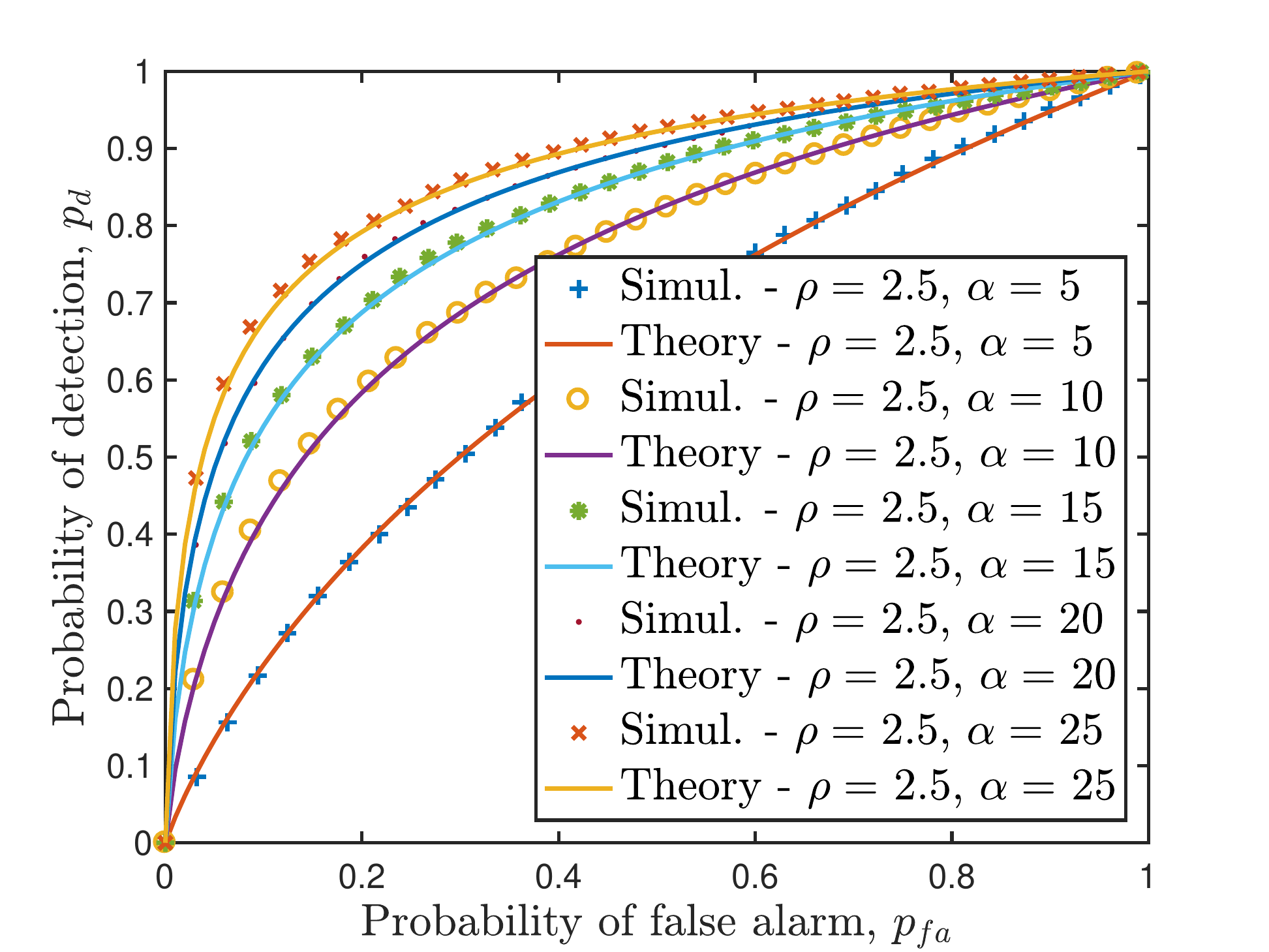}
		\caption{Theoretical and simulation results for ROC curves.}
		\label{fig:theorySim1}
	\end{subfigure}
	\caption{GLRT-CFAR for \ref{itm:1}: unknown $\alpha$ and known $h$.}
\end{figure*}
\begin{figure*}[!t]
	\centering
	\begin{subfigure}[t]{0.48\textwidth}
		\centering 
		\includegraphics[width=1\textwidth]{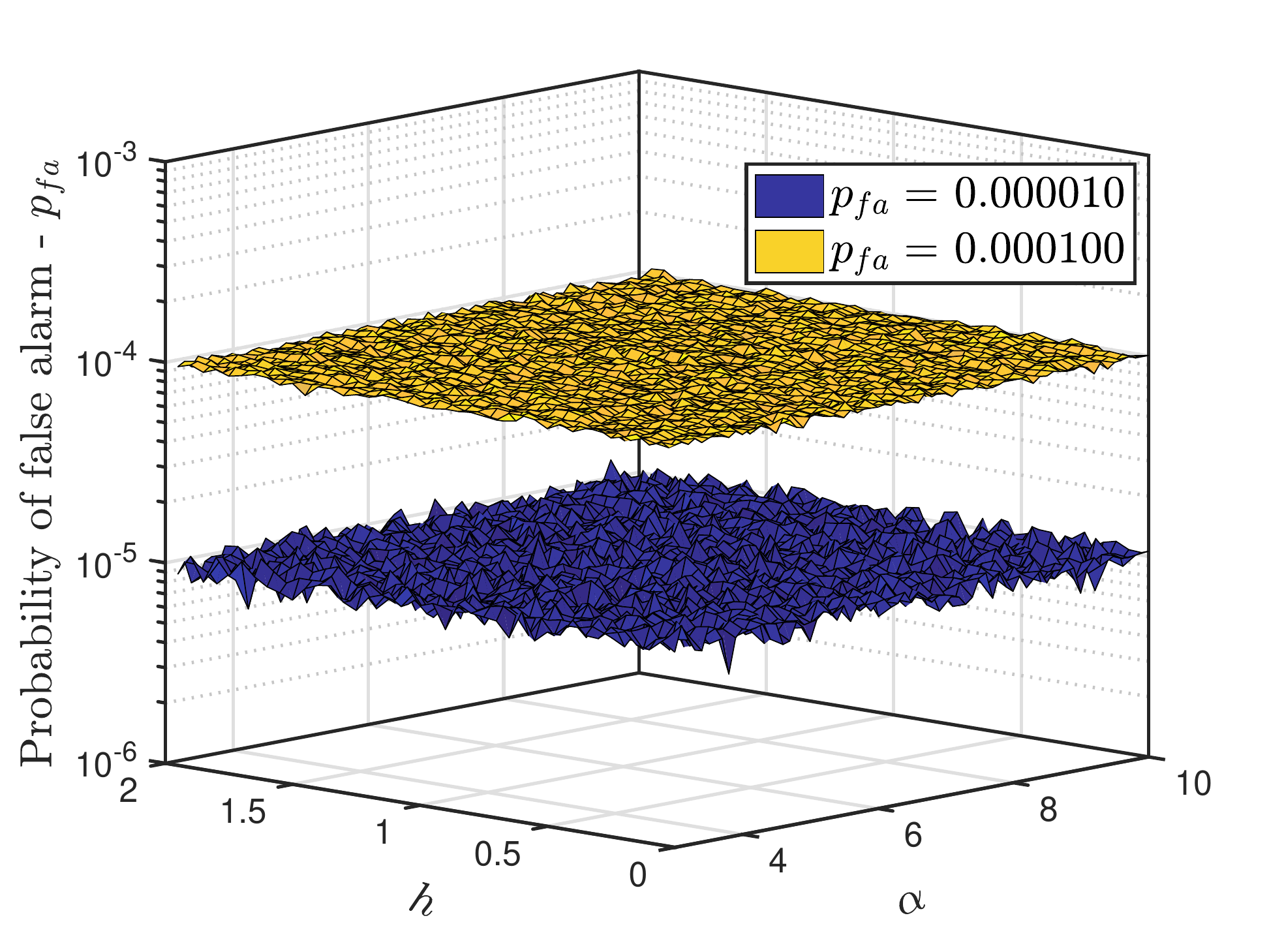}
		\caption{CFAR validation for varied $\alpha$ and $h$.}
		\label{fig:CFAR-h alpha}
	\end{subfigure}	
	\begin{subfigure}[t]{0.48\textwidth}
		\centering 
		\includegraphics[width=1\textwidth]{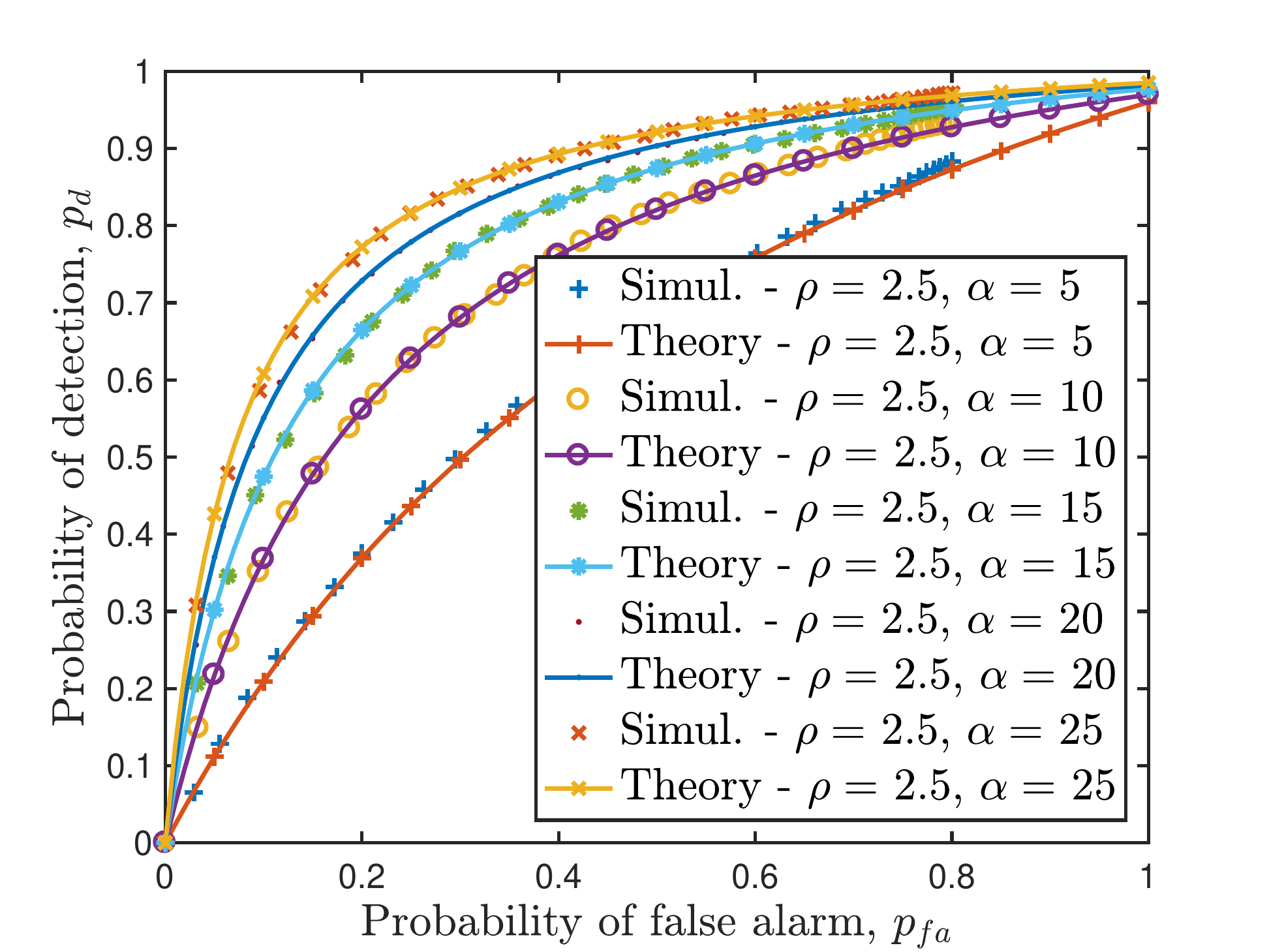}
		\caption{Theoretical and simulation results for ROC curves.}
		\label{fig:theorySim2}
	\end{subfigure}
	\caption{GLRT-CFAR for \ref{itm:2}: unknown $\alpha$ and $h$.}
\end{figure*}
	
\section{Simulation Results}
In this section, we validate the theoretical results which we derived in the previous sections with extensive Monte Carlo simulations for each case separately. We also comment on the performance of the proposed detectors with respect to the number of reference window cell observations.
\begin{figure*}[t]
	\centering
	\begin{subfigure}[t]{0.48\textwidth}
		\centering
		\includegraphics[width=1\textwidth,clip,trim=0cm 0cm 0cm 0cm]{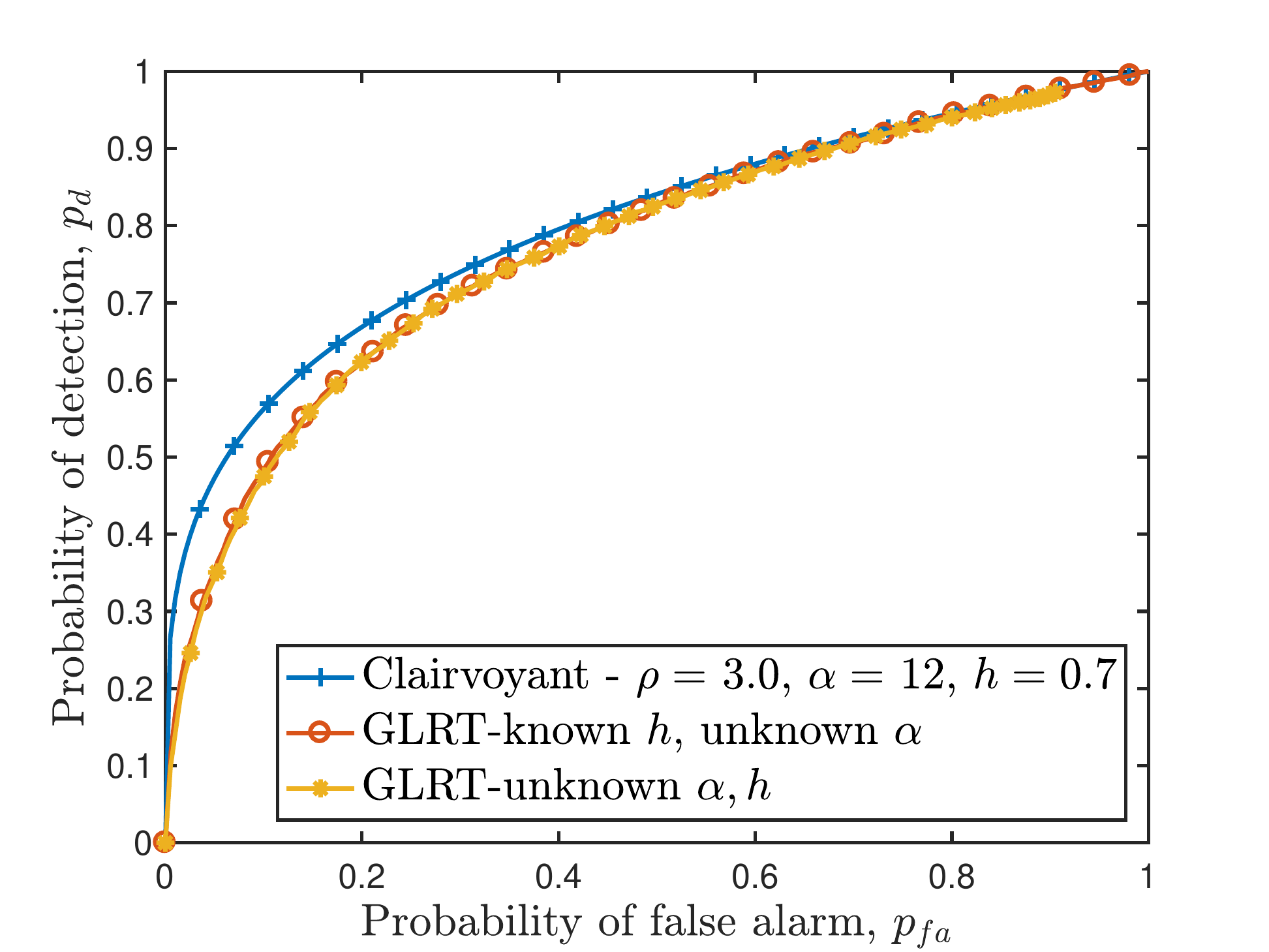}
		\caption{ROC curves when reference window obs. n=4.}
		\label{fig:comparisionN3}
	\end{subfigure}
	\begin{subfigure}[t]{0.48\textwidth}
		\centering
		\includegraphics[width=1\textwidth]{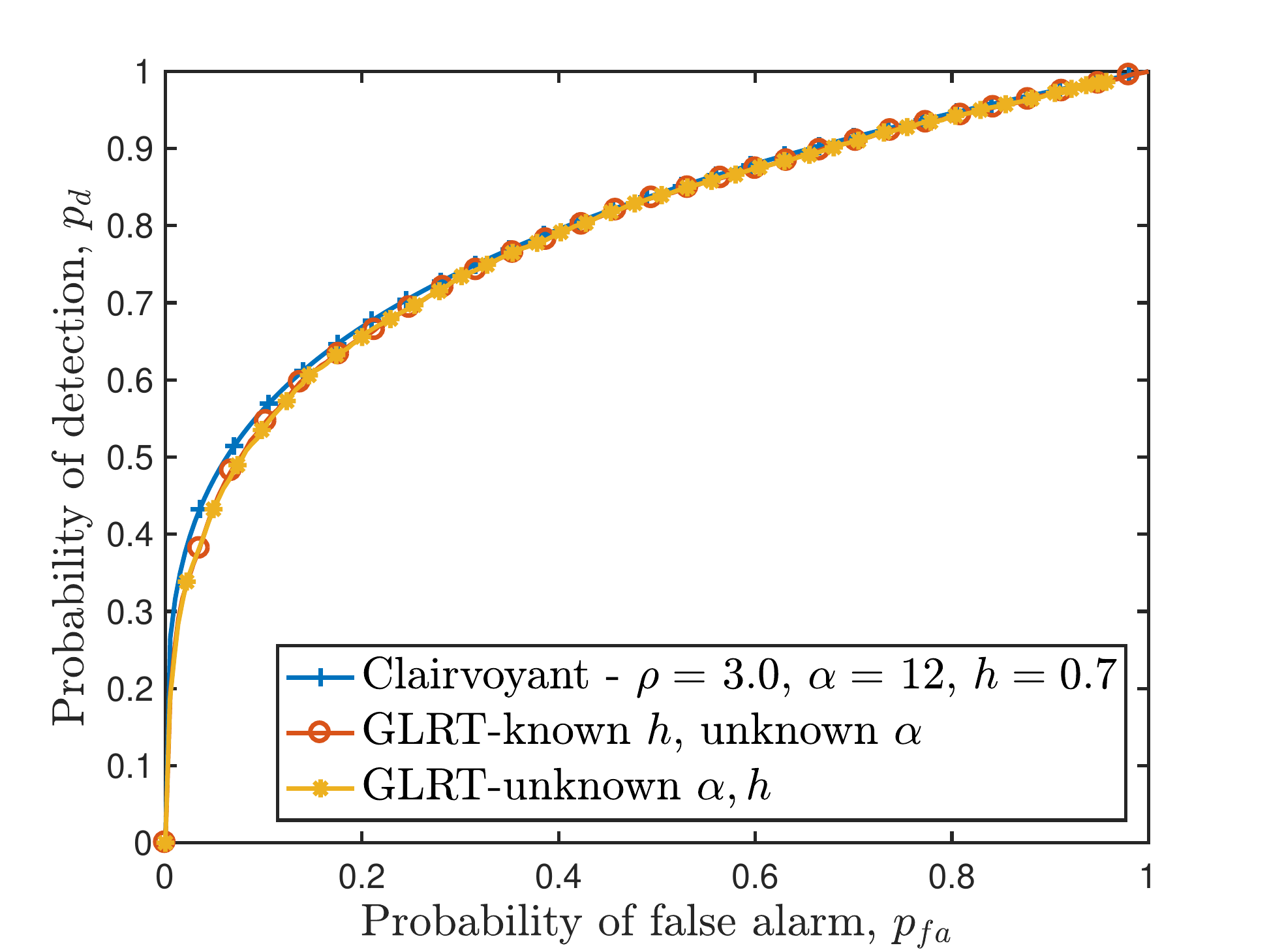}
		\caption{ROC curves when reference window obs. n=8.}
		\label{fig:comparisionN8}
	\end{subfigure}
	\caption{Comparison of GLRT-CFAR tests with respect to no. of reference window observations.}
\end{figure*}

\correct{For \ref{itm:1}, we validate the CFAR property of the test \eqref{eq:glrtStatistic} by plotting $p_{fa}$ against varied unknown clutter parameter $\alpha$  in the range of $[5,12]$. In the Fig. \ref{fig:glrt_cfar1}, we consider three levels  $P_{fa}$ taking values from $\{1, 6, 10\}\times 10^{-5} $, and clearly, $p_{fa}$ (i.e., simulated) remains constant across the range of $\alpha$, thus validating CFAR. In the simulation, we considered $10^8$ Monte Carlo runs, and by considering more data runs one gets even flatter or constant $p_{fa}$. Further, we also validate the  $p_{fa}$ \eqref{eq:threshold-pfa1}, $p_d$ \eqref{eq:pd1} expressions with the Monte Carlo simulations by plotting ROC curves  Fig. \ref{fig:theorySim1}. We varied $\alpha \in [5,20]$ in steps of $0.05$, $\rho=2.5$ and $h=0.7$, and clearly, the ROC curves depict that the theoretical and simulation results are in good accord.
	
	For \ref{itm:2}, we validate the CFAR property of the test \eqref{eq:glrt2} by plotting $p_{fa}$ against varying both the unknown clutter parameters $\alpha \in [5,12], \text{ and } h \in [0.5,2]$ as shown in the Fig. \eqref{fig:CFAR-h alpha}. Here we consider two levels of $p_{fa}=\{1, 10\}\times 10^{-5} $, and clearly, we see $p_{fa}$ (i.e., simulated) remaining constant across the range of $\alpha \text{ and } h$ thus validating CFAR. In the simulation, we considered $3\times10^7$ Monte Carlo runs, and by considering more runs, one gets even precise and constant $p_{fa}$. Further, we also validate the  $p_{fa}$ \eqref{eq:threshold-pfa2}, $p_d$ \eqref{eq:pd2} expressions with Monte Carlo simulations by plotting ROC curves in Fig. \ref{fig:theorySim2}. We varied $\alpha \in [5,20]$ in steps of $0.1$, $h \in [0.001, 2]$ in steps of $0.05$, and $\rho=2.5$, and clearly, the ROC curves depict that the theoretical and simulation results match closely.}

In reality, we face mainly composite vs. composite hypothesis testing as the assumption of accurate knowledge of clutter parameters is impractical. The popular strategy is to estimate the unknown parameters and then apply NP lemma or simply substituting MLEs of the unknown parameters in the LRs as in GLRT. As the estimates are from limited observations, GLRT is sub-optimal as one cannot attain the true values of the unknown parameters, as observed in Fig. \ref{fig:comparisionN3} for $n=4$. On the other hand, when there are more observations from reference window cells, the estimates are better, and the detection performance approaches the upper bound as depicted in Fig. \ref{fig:comparisionN8} for $n=8$. \correct{However, we can’t increase the reference window cells indefinitely as the very objective of adaptive thresholding to the changes in homogeneous clutter is not met \cite{barton_radar_2004}. In other words, such a CFAR detector wouldn't capture the varying trends in the homogeneous clutter. In such scenarios, CFAR detector performance loss is analyzed using clairvoyant upper bounds given in Fig. \ref{fig:FamilyOfRoc} for different parameters.} 
	\section{Conclusion}	
In this paper, we formulated the aircraft detection problem as a two-sample Pareto vs. Pareto composite hypothesis test for comparing tail indices with scale as the nuisance parameter. We then solved it systematically by GLRT approach for the following cases  
	\begin{enumerate}[label=\textnormal{case (\alph*)},align=left]
	\item \hspace{-4pt}\textnormal{:} when $\alpha$ is unknown and $h$ is known;
	\item \hspace{-4pt}\textnormal{:} when both $\alpha$ and $h$ are unknown.
	\end{enumerate}
In both cases, we derived the test statistic and arrived at the $p_{fa}, p_d$ expressions. We then validated our results with the extensive Monte Carlo simulations. We further showed that the test so obtained has CFAR property, which we also validated via simulations. In the future, we plan to study the  robustness of the proposed detectors under different realistic scenarios, such as in the presence of clutter edges and interfering targets. 
\newpage
	\appendix
	Simplification of equation \eqref{eq:LR} follows from \eqref{eq:appendixSimply}.
		\begin{align}\label{eq:appendixSimply}
		\lambda(\bo{x},y)&=\frac{\hat{\alpha}_{\Theta_0}^{n+1} h^{{\hat{\alpha}_{\Theta_0}}(n+1)} y^{-(\hat{\alpha}_{\Theta_0}+1)} {\left(\prod_{i=1}^{n} x_i\right)^{-(\hat{\alpha}_{\Theta_0} +1)}} }
		{\hat{\rho}_{\Theta} \hat{\alpha}_{\Theta}^{n}h^{(\hat{\alpha}_{\Theta}n+\hat{\rho}_{\Theta})} y^{-(\hat{\rho}_{\Theta}+1)}     {\left(\prod_{i=1}^{n} x_i\right)^{-(\hat{\alpha}_{\Theta} +1)} }        } \\
		&=\frac{\left[\frac{n+1}{\Lambda(\bo{x})+\Lambda(y)}\right]^{n+1}h^{\left(\frac{(n+1)^2}{{\Lambda(\bo{x})+\Lambda(y)}}\right)}    y^{-\left(\frac{n+1}{\Lambda(y)+\Lambda(\bo{x})}+1\right)}(\prod_{i=1}^{n} x_i)^{-\left(\frac{n+1}{\Lambda(y)+\Lambda(\bo{x})}+1\right)}   }{ \frac{1}{\Lambda(y)}\left[\frac{n}{\Lambda(\bo{x})}\right]^{n} h^{\left(\frac{n^2}{\Lambda(\bo{x})}+\frac{1}{\Lambda(y)}\right)}  y^{-\left( \frac{1}{\Lambda(y)}+1\right)}  \left(\prod_{i=1}^{n} x_i\right)^{-\left(\frac{n}{\Lambda(\bo{x})}+1\right)} }\\
		&=\frac{(n+1)^{n+1}{\Lambda(\mathbf{x})}^n\Lambda(y)}{n^n(\Lambda(y)+\Lambda(\mathbf{x}))^{n+1}}
		\frac{\left(\frac{y}{h}\right)^{-\left(\frac{n+1}{\Lambda(\bo{x})+\Lambda(y)}\right)}  h^{\left(\frac{n+1}{\Lambda(\bo{x})+\Lambda(y)}\right)n} (\prod_{i=1}^{n} x_i)^{-\left(\frac{n+1}{\Lambda(y)+\Lambda(\bo{x})}\right)} }{ \left(\frac{y}{h}\right)^{ -\frac{1}{\Lambda(y)}}    h^{\left(\frac{n^2}{\Lambda(\bo{x})}\right)}    \left(\prod_{i=1}^{n} x_i\right)^{-\left(\frac{n}{\Lambda(\bo{x})}\right)} }
		\\
		&=\frac{(n+1)^{n+1}}{n^n}\frac{\frac{\Lambda(y)}{\Lambda(\mathbf{x})}}{(\frac{\Lambda(y)}{\Lambda(\mathbf{x})}+1)^{n+1}}{\left(\frac{y}{h}\right)^{\left(\frac{1}{\Lambda(y)}-\frac{n+1}{\Lambda(y)+\Lambda(\mathbf{x})}\right)}\left(\frac{\prod_{i=1}^{n} x_i}{h}\right)^{\left(\frac{n}{\Lambda(\bo{x})}-\frac{n+1}{\Lambda(y)+\Lambda(\mathbf{x})}\right)}}\\
		&={(n+1)^{n+1}} \frac{n\frac{\Lambda(y)}{\Lambda(\bo{x})}}{\left((n\frac{\Lambda(y)}{\Lambda(\bo{x})} +n)\right)^{n+1} } \left(e^{\Lambda(y)}\right)^{\left(\frac{1}{\Lambda(y)}-\frac{n+1}{\Lambda(y)+\Lambda(\mathbf{x})}\right)}    
		\left(e^{\Lambda(\bo{x})}\right)^{\left(\frac{n}{\Lambda(\bo{x})}-\frac{n+1}{\Lambda(y)+\Lambda(\mathbf{x})}\right)}\\
		&={(n+1)^{n+1}} \frac{n\frac{\Lambda(y)}{\Lambda(\bo{x})}}{\left((n\frac{\Lambda(y)}{\Lambda(\bo{x})} +n)\right)^{n+1} } e^{\left( 0\right)}.
		\end{align}
	Further, from equation \eqref{eq:sizeSimplification1}, arriving at threshold$-\space p_{fa}$ relation \eqref{eq:threshold-pfa1} is as follows: 
	
	\begin{align}
	p_{fa} &=\sup_{{\alpha} \in {(0,\infty)}}\Pr\left( \frac{B}{C}>\gamma_{1} \right)\\
	&=\sup_{{\alpha} \in {(0,\infty)}}\int_{c=0}^{\infty}\int_{b=\gamma_{1} c}^{\infty}f_{BC}\left(b,c\right) \dd{}b\:\dd c\\
	&\stackrel{(a)}{=}\sup_{{\alpha} \in {(0,\infty)}}\int_{c=0}^{\infty}\left(\int_{b=\gamma_{1}^{} c}^{\infty}f_B\left(b\right)\dd b\right) f_C(c)\:\dd c\\
	&\stackrel{(b)}{=}\sup_{{\alpha} \in {(0,\infty)}}\mathbf{E}_C\left[\int_{b=\gamma_{1}^{} c}^{\infty}f_B\left(b\right)\dd b\right]\\
	&\stackrel{(c)}{=}\sup_{{\alpha} \in {(0,\infty)}}\mathbf{E}_C\left[e^{(-\alpha\gamma_{1}^{}c)}\right],\\
	&\stackrel{(d)}{=}\sup_{{\alpha} \in {(0,\infty)}}\left[1-\frac{1}{n\alpha}\left(-\alpha\gamma_{1}^{}\right)\right]^{-n}\\
	&=\left[1+\frac{\gamma_{1}^{}}{n}\right]^{-n}
	\end{align}
	where the equalities are justified as follows:
	$(a)$ for $B \text{ and } C$ are independent random variables and their joint density products down; $(b)$ by taking expectation with respect to the $C$ random variable; $(c)$ by the complementary cdf formula; $(d)$ by applying moment generating function formula for $C$, followed by simplification.
	
	For equation \eqref{eq:pd1} following the similar lines of above justification,  but under $H_1$, the simplification steps are:
	\begin{align}
	p_d&=\Pr\left( \frac{B}{C}>\gamma_{1} ;H_1\right)\\
	&=\int_{c=0}^{\infty}\int_{b=\gamma_{1} c}^{\infty}f_{BC}\left(b,c\right)\dd b\:\dd c\\
	&=\int_{c=0}^{\infty}\left(\int_{b=\gamma_{1}^{} c}^{\infty}f_B\left(b\right)\dd b\right) f_C(c)\:\dd c\\
	&=\mathbf{E}_C\left[\int_{b=\gamma_{1}^{} c}^{\infty}f_B\left(b\right)\dd b\right]\\
	&=\mathbf{E}_C\left[e^{(-\rho\gamma_{1}^{}c)}\right]\\
	&=\left(1+\frac{\rho \gamma_1}{\alpha n}\right)^{-n}.
	\end{align}
	Similarly for \eqref{itm:2}, from equation \eqref{eq:size2Simplificaiton} arriving at \eqref{eq:threshold-pfa2} is as follows:
	\begin{align}
	p_{fa}&=\sup_{{(\alpha,h)} \in \Theta_0}\Pr\left(\frac{G}{D}>\gamma\right)\\
	&\stackrel{}{=}\sup_{{(\alpha,h)} \in \Theta_0}\int_{d=0}^{\infty}\int_{g=\gamma d}^{\infty}f_{GD}\:\dd g\:\dd d\\
	&\stackrel{(a)}{=}\sup_{{(\alpha,h)} \in \Theta_0}\int_{d=0}^{\infty}\left(\int_{g=\gamma d}^{\infty}f_G\left(g\right)\:\dd g\right) f_D(d)\:\dd d\\
	& \stackrel{(b)}{=}\sup_{{(\alpha,h)} \in \Theta_0} \int_{d=0}^{\infty} f_D\left(d\right)\int_{g=\gamma d}^{\infty}\frac{n}{n+1}\alpha e^{-\alpha g}\:\dd g\: \dd d,\\
	& \stackrel{}{=}\sup_{{(\alpha,h)} \in \Theta_0}\frac{n}{n+1}\int_{d=0}^{\infty}\left[{-e^{-\alpha g}}\right]^{\infty}_{g=\gamma d}f_D\left(d\right)\dd d\\
	&\stackrel{(c)}{=}\sup_{{(\alpha,h)} \in \Theta_0}\frac{n}{n+1} \mathbf{E}_D \left[e^{-\alpha\gamma d}\right]\\	
	&\stackrel{(d)}{=}\sup_{{(\alpha,h)} \in \Theta_0}\frac{n}{n+1}\left(1-\frac{1}{n\alpha}(-\alpha \gamma)\right)^{-(n-1)}\\
	&=\frac{n}{n+1}\left[1+\frac{\gamma}{n}\right]^{-(n-1)}
	\end{align}
	where the equalities are justified as follows: $(a)$ for $G \text{ and } D$ are independent random variables; $(b)$ substituting the density function for $g>0$ in \eqref{eq:densityG} as $\gamma d >0$; $(c)$  by taking expectation with respect to the $D$ random variable; $(d)$ by applying moment generating function formula for $D$, followed by simplification.
	
	\bibliographystyle{IEEEtran}
	\bibliography{try}
	
	%
	
\end{document}